\documentclass[9pt,twocolumn,twoside]{opticajnl}
\journal{opticajournal} 

\setboolean{shortarticle}{true}

\usepackage{pdfpages}

\usepackage{lineno}

\title{Vortex-comb spectroscopy}

\author[1,*]{Santeri Larnimaa}
\author[1,2,3]{Markku Vainio}

\affil[1]{Department of Chemistry, University of Helsinki, FI-00560, Helsinki, Finland}
\affil[2]{Photonics Laboratory, Physics Unit, Tampere University, Tampere, FI-33101, Finland}
\affil[3]{markku.vainio@helsinki.fi}

\affil[*]{santeri.larnimaa@helsinki.fi}

\begin{abstract}
We propose a new  Fourier-transform spectroscopy technique based on the rotational Doppler effect. The technique offers an application for optical vortex frequency combs, where each frequency component carries a unique amount of orbital angular momentum (OAM). Here, we emulate a vortex comb using a tunable single frequency laser and a collection of spiral phase plates, generating up to eleven distinct OAM  modes. Unlike in traditional Fourier-transform spectroscopy based on the Michelson interferometer (linear Doppler effect), the spectral resolution of vortex-comb spectroscopy is not limited by the mechanical scan distance of the instrument but only by the measurement time. Although the spectrometer requires just one free-running frequency comb, the down-conversion scheme resembles dual-comb spectroscopy, leading to fast mode-resolved measurements.
\end{abstract}

\setboolean{displaycopyright}{false} 

\begin{document}
\maketitle
Light carrying orbital angular momentum (OAM) is a rapidly expanding research field with applications ranging from optical communications and quantum optics to microscopy and imaging \cite{OAMreviewShen}. OAM light, also called twisted or optical vortex light is characterized by a helical wavefront $\exp(\mathrm{i}\theta l)$, where the so-called topological charge $l$ denotes that the beam carries $l\hbar$ of orbital angular momentum per photon. OAM light can  be easily generated, for example, by using cylindrical lens converters, spatial light modulators, or spiral phase plates \cite{OAMreviewWang}.\par
An intriguing phenomenon related to OAM light is the rotational Doppler effect \cite{RDEreview}. In analogy to the more familiar linear Doppler effect where light incident on a moving target can experience a frequency shift proportional to the change of linear momentum upon reflection, a  frequency shift can ensue when light interacts with a rotating target if the topological charge changes in the process. The rotational Doppler effect has gained particular interest due to the possibility of using it for ranging the rotational movement of far-away objects \cite{lavery,OAMinterferometry,OAMdualcomb}, or to characterize the modal composition of OAM light itself \cite{OAManalyser}. We propose that the rotational Doppler effect can also be used for spectroscopy, for example, to measure molecular absorption. We call the proposed technique vortex-comb spectroscopy, and it can be viewed as a type of Fourier-transform spectroscopy (FTS).

In traditional Fourier-transform spectroscopy \cite{Griffiths}, a light beam is split into two arms of an interferometer: while one of the split beams serves as a reference, the other reflects from a translating mirror that shifts each frequency component in the beam by a unique amount due to the linear Doppler effect. The two beams are combined again, sent through a sample, and collected on a detector to record the time domain signal, also called the interferogram. The interferogram arises from beating of the Doppler shifted beam against the reference beam with the result that the optical frequencies oscillating too fast to be measured 
 directly are down-converted to something slower but easily detectable intensity modulation. The Fourier transform of the interferogram reveals the spectrum of the original light source modified by the absorption of the sample, allowing  analysis of the sample composition.

Instead of linear motion, vortex-comb spectroscopy relies on rotational motion with the benefit that the resolution of the measurement is not limited by the mechanical scan distance of the instrument but only by the measurement time. We also show that the down-conversion scheme resembles dual-comb spectroscopy \cite{dualcomb} or phase-controlled Fourier-transform spectroscopy \cite{Hashimoto,Larnimaa}, which leads to  fundamentally faster measurements than with traditional FTS. A prerequisite for the technique is a light source where each frequency component carries a unique amount of OAM. As an example, optical vortex combs have been recently demonstrated \cite{liu2022integrated,chen2024integrated,LiEtAl}. We currently do not have access to such light sources, for which reason we emulate a vortex comb using a single frequency (but tunable) laser and a collection of spiral phase plates and their combinations to generate up to 11  OAM modes. As a proof of concept, we construct an experimental spectrum of an acetylene absorption feature at the 1 µm wavelength region from spectra separately measured with each of these 11 OAM modes. In the following, we first explain the down-conversion theory, after which we present our experimental setup and discuss the experimental results. 

The rotational Doppler effect can ensue upon interaction between light and various different rotating objects \cite{RDEreview}. Here, we focus on a rotating surface whose reflectivity has spatial variability. Following the formulation of Ref. \cite{derivation}, when an OAM-carrying  light beam $B(r)\exp(-\mathrm{i}2\pi\nu t)\exp(\mathrm{i}l\theta)$  is incident on such a surface, it is reflected to a collection of OAM modes $\sum_mB(r) A_{m-l}(r)\exp(-\mathrm{i}2\pi\nu t)\exp(-\mathrm{i}2\pi[m-l]f_\mathrm{rot}t)\exp(\mathrm{i}m\theta)$, each having experienced a frequency shift $[m-l]f_\mathrm{rot}$ proportional to the rotational frequency of the reflective surface and to the change of the topological charge. Here, $B(r)$ and $A_{m-l}(r)$ are the complex amplitudes of the incident and reflected fields, respectively. One way of detecting the rotational Doppler shift is the so-called fringe method \cite{fringe}, where  the OAM-carrying field and its OAM mode inverted ($l\rightarrow-l$) copy are jointly incident on the rotating surface. The inverted field similarly produces a collection of modes $n$, and for each mode $m$ one can find a common detection mode for which $m=n$. While one of the modes experiences a red shift with respect to the incident light, the other undergoes a blue shift, leading to intensity modulation at the frequency $([n+l]-[m-l])f_\mathrm{rot}=2lf_\mathrm{rot}$ when the two reflected fields are collected and interfere on a photodetector. A more detailed derivation is shown in Supplement 1.

\begin{figure}[ht]
\centering
\includegraphics[width=1\linewidth]{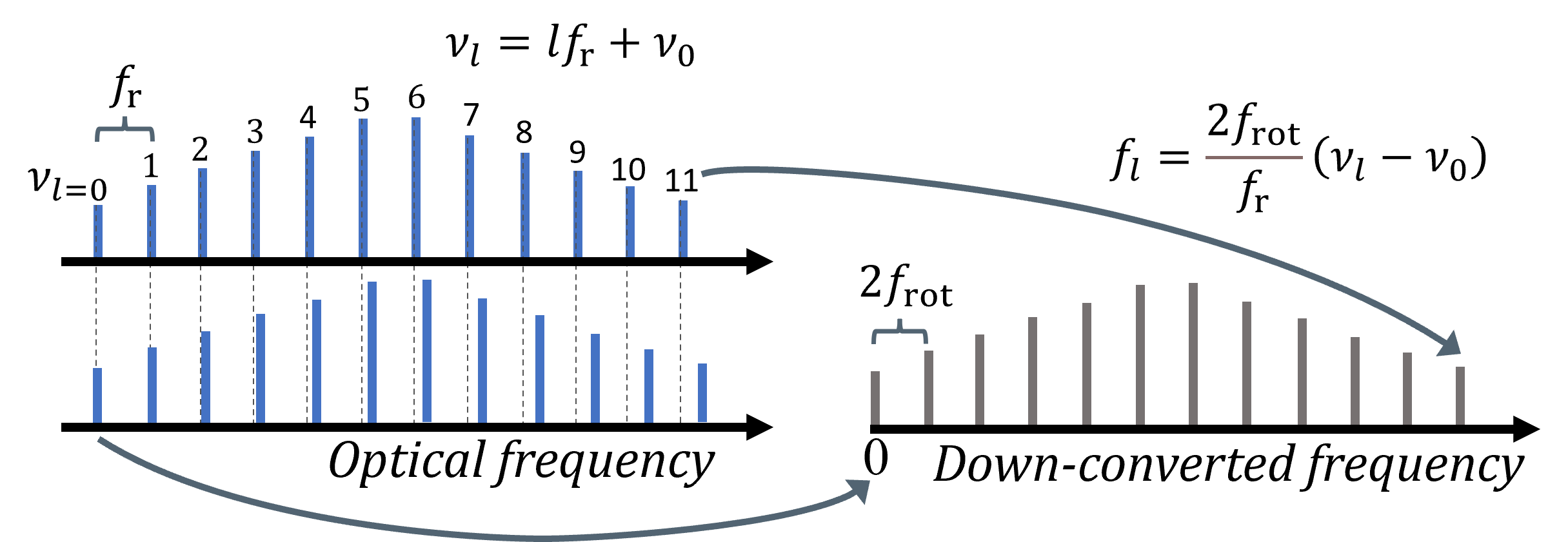}
\caption{\justifying Illustration of the down conversion in vortex-comb spectroscopy. Owing to the rotational Doppler effect, each mode $l$ of the vortex comb is frequency-shifted by a unique amount, resulting in a down-converted  spectrum with $2f_\mathrm{rot}$ line spacing, where $f_\mathrm{rot}$ is the rotation frequency of the reflective target that induces the rotational Doppler shift.}
\label{fig:downConversion}
\end{figure}

Imagine that the original light field is an optical frequency comb with a frequency spacing of $f_\mathrm{r}$ and that each of the comb lines carries a unique amount of OAM. Then, each optical frequency will be down-converted to a unique  frequency such that the down-converted comb has a line spacing of $2f_\mathrm{rot}$, as exemplified in Fig. \ref{fig:downConversion}. If the optical frequency-to-OAM mode  mapping is $\nu_l=lf_\mathrm{r}+\nu_0$, we arrive at the down-conversion relation
\begin{equation}\label{eq1}
f_l=2lf_\mathrm{rot}=\frac{2f_\mathrm{rot}}{f_\mathrm{r}}  (\nu_l-\nu_0 )\,,\end{equation}
where $2f_\mathrm{rot}/f_\mathrm{r}$ is the down-conversion factor and $\nu_0$ indicates the optical frequency that has $l=0$ and is thus mapped to the zero frequency. Note that this down-conversion relation is of the same form as in dual-comb spectroscopy \cite{dualcomb} or in phase-controlled Fourier-transform spectroscopy \cite{Hashimoto}, which leads to faster measurements than in traditional Fourier-transform spectroscopy as only the optical bandwidth of interest is down-converted to efficiently fill the detection bandwidth \cite{Larnimaa}. Note further that the effective ''repetition rate difference'' between the two interfering combs is $2f_\mathrm{rot}$. This means that mode-resolved measurements can be attained in a measurement time of $1/(2f_\mathrm{rot})$.

\begin{figure}[ht]
\centering
\includegraphics[width=\linewidth]{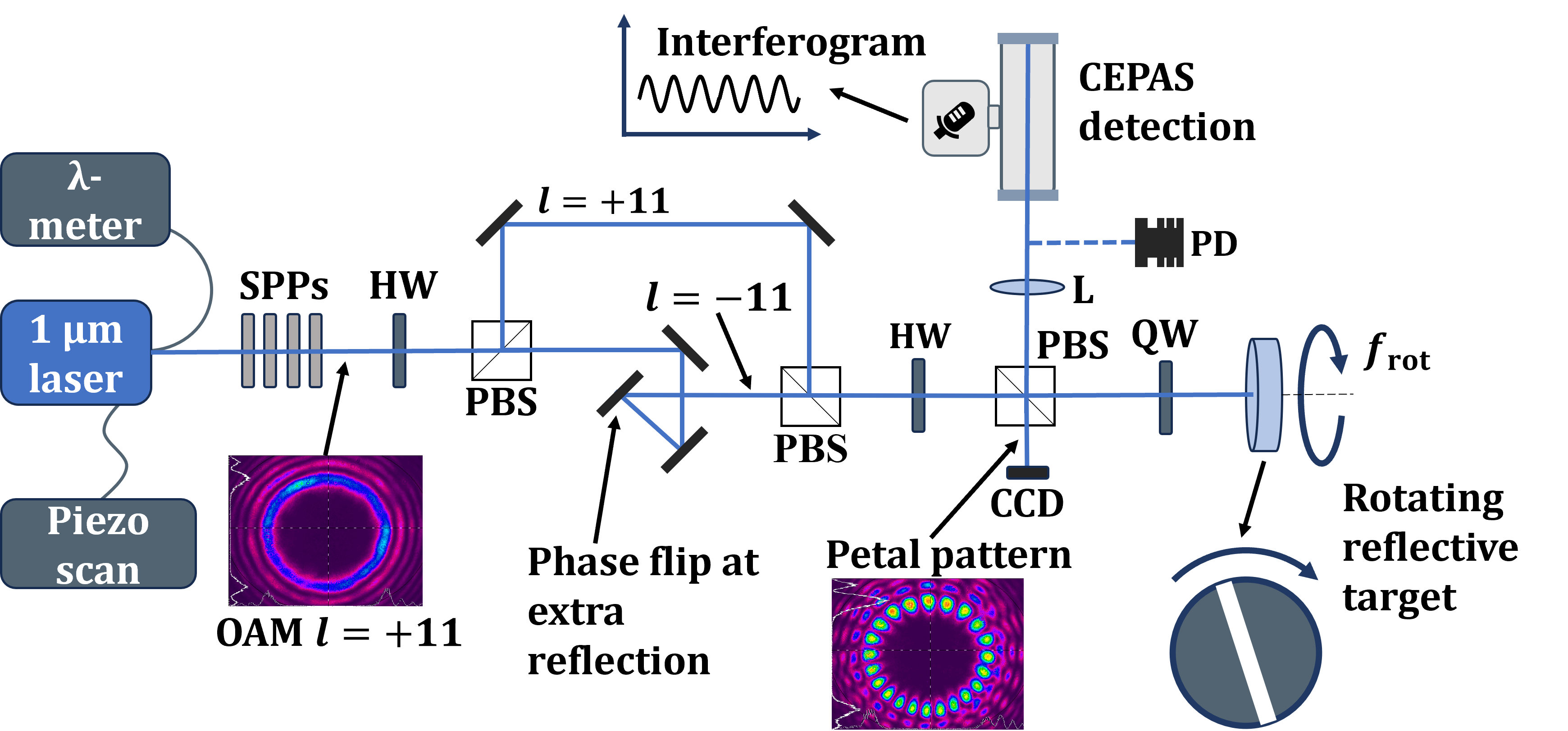}
\caption{\justifying Experimental setup illustrating a measurement with the OAM mode $l=11$. SPP: spiral phase plate, HW: half-wave plate, PBS: polarizing beam splitter, QW: quarter-wave plate, L: lens, CEPAS: cantilever-enhanced photoacoustic detector filled with gaseous acetylene for the spectroscopic measurements, PD: photodetector for general characterization measurements, $f_\mathrm{rot}$: the rotational frequency of the target. The insets illustrating the OAM $l=11$ beam and the corresponding petal pattern have been measured with a CCD camera.}
\label{fig:experimental}
\end{figure}

To demonstrate molecular spectroscopy with an emulated vortex comb, we apply the fringe method for detection and use the implementation shown in  Fig. \ref{fig:experimental}. Light from a tunable 1 µm fiber laser (NKT Koheras Basik) is collimated and sent through a collection of spiral phase plates (Vortex Photonics). 
We use a total of four plates with the topological charges of 1, 1, 3 and 6 in different combinations to yield OAM modes 1, 2, ..., and 11. The beam carrying a chosen topological charge (spiral phase plate combination) is split into two arms and combined again. In one of the arms, an extra reflection flips the phase inverting the OAM mode \cite{fringe}. Therefore, upon combining the beams a petal pattern is formed. The petal pattern is incident on a rotating mirror onto whose surface a suitable target cut out of paper is taped. The rotation is achieved by attaching the mirror onto an optical chopper (MC1F10HP high precision blade, Thorlabs MC2000B Optical Chopper System) by using a custom mount with tilt compensation. The tilt compensation ensures that the mirror surface normal is parallel to the rotation axis. In addition, the chopper head has an xy-adjustment option to center the rotation axis onto the beam propagation axis.  Finally, the reflected light from the rotating mirror is gathered and focused for detection.

The specific shape of the target or its positioning on the rotating surface does not affect the frequency of the down-converted signal. However, it does affect its amplitude. As a rule of thumb, for a petal pattern with $2l$-fold rotational symmetry, the signal is maximized using a target of the same symmetry \cite{fringe}. Instead of maximizing the signal for a specific mode, we want to ensure  similar modulation amplitudes for all the OAM modes. This is achieved at least with a target with 2-fold symmetry such as a reflective rectangle, provided that the rectangle is narrow enough. As the modulation amplitude decreases as a function of the topological charge $l$, it is beneficial to optimize the target for the largest $l$ value in use. The target we use is a 0.3 mm wide rectangular slit cut out of paper and taped onto the rotating mirror surface. The width corresponds to half of the distance between adjacent intensity maxima in the petal pattern for the OAM $l=11$ mode (the diameter of the petal pattern on the mirror was 4.1 mm). Experimental characterization showed that the modulation amplitude for the $l=11$ mode was 16 \% of that of the $l=1$ mode.

\begin{figure}[ht]
\centering
\includegraphics[width=\linewidth]{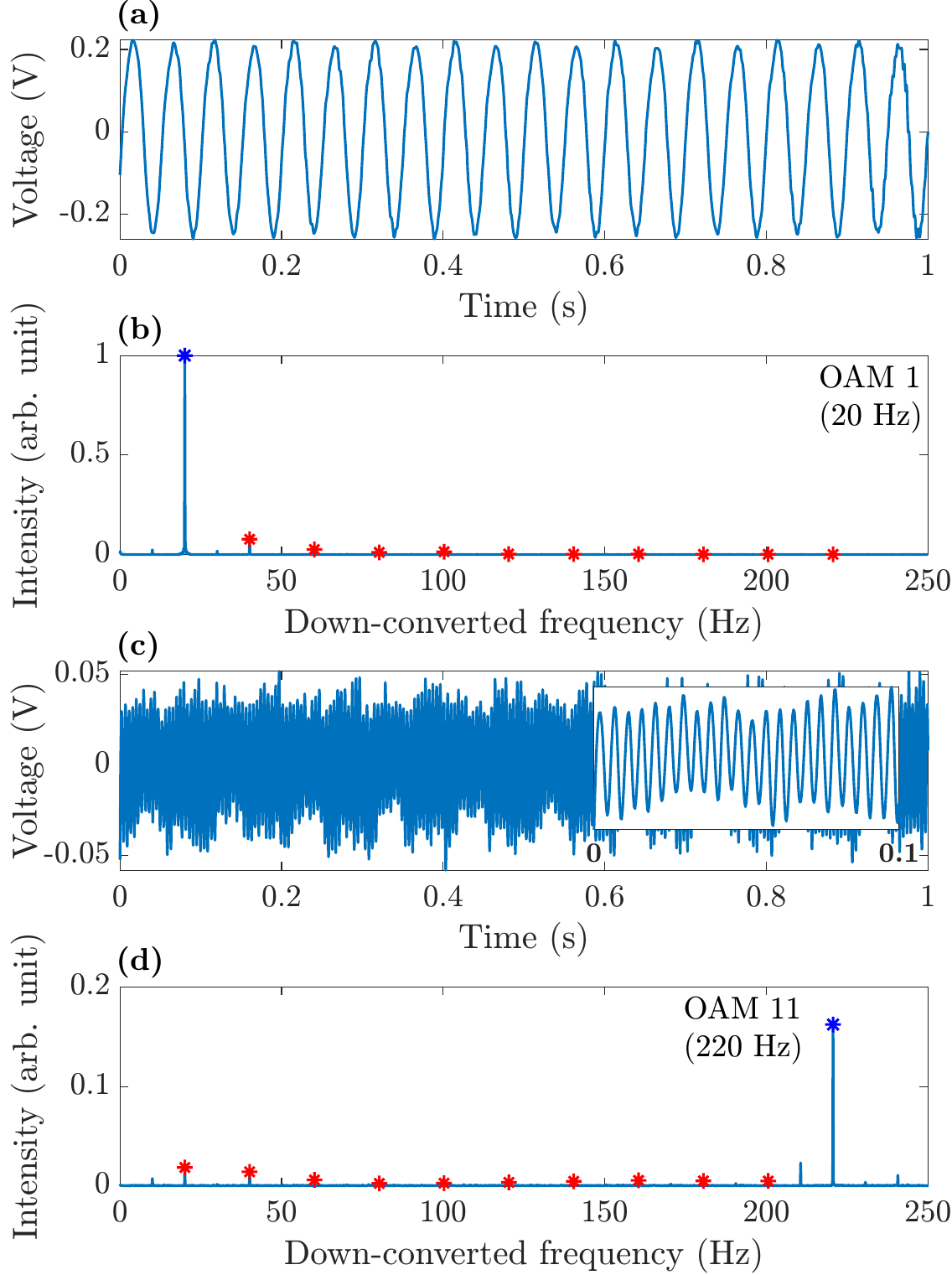}
\caption{\justifying Typical interferograms and their Fourier transforms measured without an absorbing gas with the OAM modes $l=1$ (panels a and b) and $l=11$ (panels c and d) with rotational frequency $f_\mathrm{rot}  = 10$ Hz. The total length of the time window was 12 s, but here only 1 s is shown (0.1 s for the inset in panel c). The main spectral peak at $2lf_\mathrm{rot}$ is marked with a blue asterisk. The unwanted side peaks at even multiples of $f_\mathrm{rot}$ have been marked with red asterisks. The largest side peak is 8 \% (12 \%) of the corresponding main peak for the OAM 1 (OAM 11) measurement.}
\label{fig:ifrgs}
\end{figure}

Fig. \ref{fig:ifrgs} shows examples of typical time domain signals and their Fourier-transforms recorded with a photodetector (Thorlabs PDA20CS-EC) and digitized with an oscilloscope (Keysight DSO1052B). The target rotation frequency was 10 Hz.  It can be seen that for each OAM mode the largest modulation peak appears at the expected frequency ($2lf_\mathrm{rot}$). However, in addition to the main peak, side peaks appear. Those additional peaks  at even multiples of $f_\mathrm{rot}$  would lead to crosstalk between different OAM modes and distort the measured absorption features in the case of using a vortex comb light source; in our case the crosstalk has no effect as the measurements were performed one OAM mode at a time and the final absorption spectrum is constructed using the main peak signals only. The side peaks are due to imperfect overlap of the combined beams or due to nonideal centering of the spiral phase plates leading to degradation of the petal pattern. In addition, imperfect centering of the rotation axis on the beam propagation axis  creates unwanted side modes \cite{misalignment}. It is an open question whether the situation would improve using an optical vortex comb light source where all the OAM modes are emitted from the same device, instead of having to remove and reattach the phase plates as is done here, which possibly affects the alignment. Other research groups  have used spatial light modulators to both generate the OAM modes and to simulate the rotating target to enable precise alignment control \cite{lavery,derivation,fringe,OAManalyser,OAMdualcomb}. 

Our setup was constructed for the 1 µm wavelength region, where strong absorption by molecules is scarce. One of the strongest absorbing molecules available in the HITRAN database \cite{HITRAN} is acetylene, whose rovibrational line R17e at 9402.16 cm$^{-1}$ (1063.59 µm) belonging to the $\nu_1 \nu_2 \nu_3 \nu_4 \nu_5 l_4 l_5=2100101$ combination band 
was measured in this work. The line intensity is $2.3\times10^{-25}\,\mathrm{cm}^{-1}/\mathrm{cm}^{-2}$, corresponding to $1.3\times10^{-4}$ peak absorbance for pure acetylene at 270 mbar and 10 cm absorption path length. To measure this weak absorption, we employed cantilever-enhanced photoacoustic spectroscopy (CEPAS), a highly sensitive spectroscopic detection technique \cite{PASreview,Tomberg} where  modulation of the  optical power (here, due to the rotational Doppler effect) is converted to sound waves by the absorption inside the gas cell and detected by a cantilever microphone. Basically, the acetylene-filled CEPAS cell (Gasera PA201) is simply a means to record the interferograms \cite{Larnimaa}. To ensure high enough signal-to-noise ratio (SNR), the optical power of the 1 µm laser was amplified (IPG YLR-20-1064-LP-SF) such that approximately 40 mW was incident on the rotating target. It is noteworthy that the target will block a portion of this power and introduce the OAM mode-dependent  modulation to the remaining power \cite{fringe}. The high optical power  is not a prerequisite for vortex-comb spectroscopy per se but is required here due to the weak molecular absorption; CEPAS signal is proportional to optical power.

\begin{figure*}
\centering
\includegraphics[width=\linewidth]{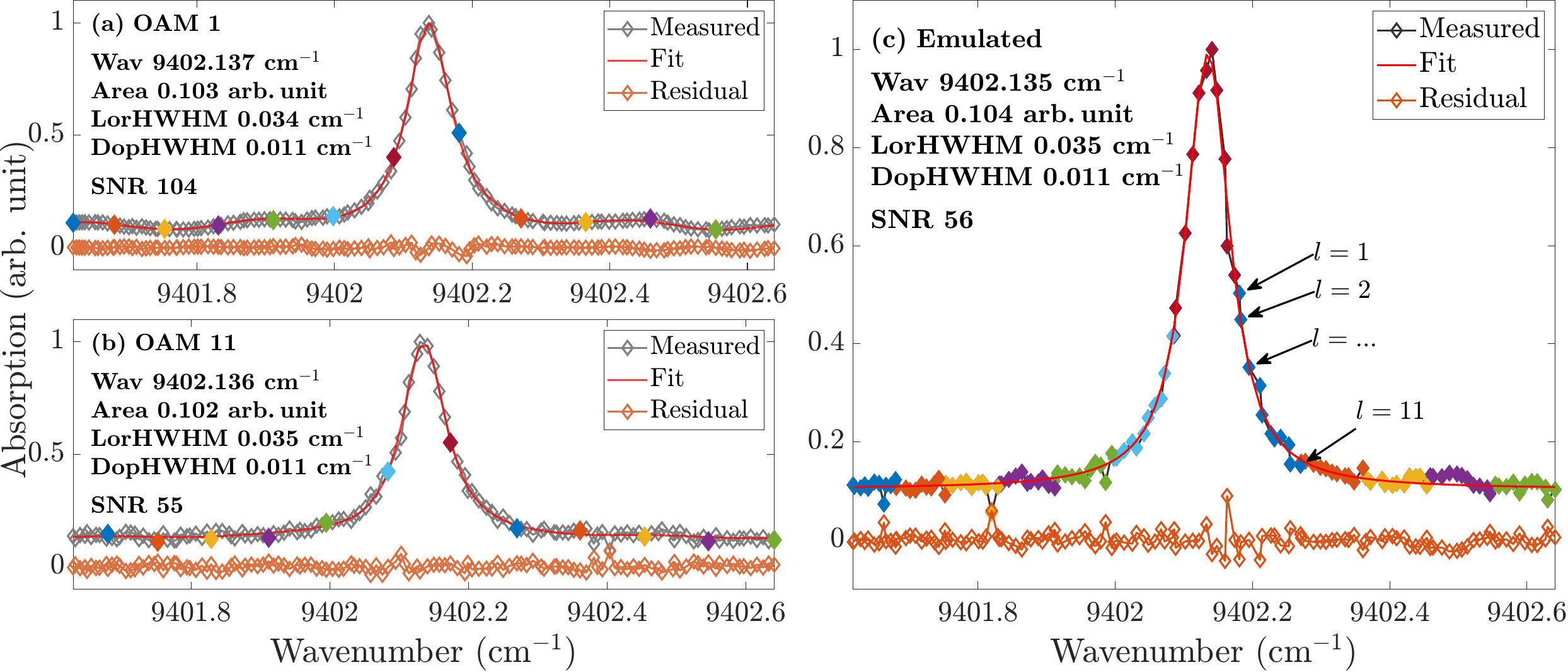}
\caption{\justifying Example acetylene spectra measured with the OAM mode $l=1$ (a) and  $l=11$  (b). Panel (c) shows an experimental spectrum constructed from points chosen from each spectrum measured with the different OAM modes. The filled symbols in (a) and (b) indicate that they were chosen for the construction. The colour coding in (c) illustrates 12 subsequent measurements with the emulated vortex comb consisting of 11 OAM modes: each colour indicates a single measurement with the comb after which its center optical frequency is tuned to the next measurement window, indicated by another colour. The legend shows information on a Voigt profile fit, such as the line center and the Lorentzian and Doppler halfwidths. The signal-to-noise ratio (SNR) is defined as the ratio of the fitted peak height to the standard deviation of the residuals. }
\label{fig:spectra}
\end{figure*}

Panels (a) and (b) of Fig. \ref{fig:spectra} show typical molecular absorption spectra measured with the separate OAM modes. For a chosen OAM mode, the measurement was performed by scanning the laser wavenumber in steps, while the laser wavenumber was monitored using a wavelength meter (EXFO WA-1500-NIR-89). Three 2.4 s long interferograms were measured, apodized with a sine lobe, Fourier-transformed, and averaged at each scan step. For each averaged spectrum, the main peak height was read and stored. The spectra shown in Fig. \ref{fig:spectra} (a) and (b)  
 were then obtained by plotting the corresponding main peak value as a function of the laser wavenumber.

Panel (c) of Fig. \ref{fig:spectra} shows the emulated vortex-comb  spectroscopy measurement, constructed using points from every spectrum measured with the different OAM modes: every 12th point starting from the 1st point in the OAM 1 spectrum, every 12th point starting from the 2nd point in the OAM 2 spectrum, and so on. More information on the signal construction is given in Supplement 1. The spectral point spacing in the resulting emulated vortex-comb measurement is 0.0078 cm$^{-1}$ (230 MHz). This corresponds to 12 subsequent measurements using an optical vortex comb spanning 0.085 cm$^{-1}$ (25.5 GHz; 11 OAM modes) such that the center optical frequency of the comb is tuned from one measurement window to the next. The resulting spectrum is of excellent quality, and a Voigt line profile fit (with the Doppler half width-at-half-maximum fixed to its theoretical value) reveals a Lorentzian HWHM of 0.035 cm$^{-1}$. This value implies a measurement pressure of 270 mbar (acetylene self-broadening for this absorption line is 0.129 cm$^{-1}$/atm
), which matches well with the pressure inside the cell measured with an external pressure sensor (Honeywell HSCMRNN015PAAA5).

In conclusion, we have demonstrated that the rotational Doppler effect can be used for spectroscopy. The proposed technique offers an interesting new application for optical vortex combs with certain benefits. For example, the down-conversion resembles dual-comb spectroscopy, but due to the auto-correlation nature of the technique, the experiment is much simplified as there is no need for two mutually coherent light sources. Compared to traditional Fourier-transform spectroscopy, the setup can be compact even in high resolution measurements as the resolution is not limited by the scan distance of the instrument but only by the measurement time. Here, mode-resolved measurements could be attained in a measurement  time of $1/(2f_\mathrm{rot})=50\,\mathrm{ms}$. With the 230 MHz comb-line spacing of the emulated comb, such resolution would require a mechanical scan length greater than 0.65 m in traditional FTS based on the Michelson interferometer \cite{Griffiths}.

The spectra that we measured are of high quality despite the weak absorption at the 1 µm  wavelength region. To eliminate the added layer of complexity from having to use CEPAS detection, it would be beneficial to transfer the setup, for example, to the 1.5 µm wavelength region where the typical absorption features are five orders of magnitude stronger. In addition, the technique should be tested using an actual optical vortex comb light source and not an emulated one. In particular, further analysis is needed to study and minimize the comb-mode crosstalk that can  emerge from imperfect optical alignment.

\begin{backmatter}

\bmsection{Funding} University of Helsinki.

\bmsection{Acknowledgments} We thank Dr. Markus Metsälä, Dr. Juho Karhu, and Mikhail Roiz for feedback on the project. S. Larnimaa acknowledges financial support from the CHEMS doctoral program of the University of Helsinki.

Author contributions: M. Vainio and S. Larnimaa conceptualized the project. S. Larnimaa constructed the experimental setup, performed the measurements, analysed the data, and wrote the manuscript; M. Vainio supervised the project, and reviewed the manuscript.

\bmsection{Disclosures} The authors declare no conflicts of interest.

\bmsection{Data availability} Data underlying the results presented in this paper are not publicly available at this time but may be obtained from the authors upon reasonable request.

\bmsection{Supplemental document}
See Supplement 1 for supporting content. 

\end{backmatter}%

\clearpage
{

\section*{Supplementary material}

# VORTEX-COMB SPECTROSCOPY: SUPPLEMENTAL DOCUMENT

## 1. The rotational Doppler effect

Following the formulation of Ref. [1], here we derive the intensity modulation signal that emerges from having an optical vortex comb light source (and its OAM mode inverted counterpart) incident on a rotating surface. For succinctness, let's assume a vortex beam with only two optical angular frequencies $\omega_1$ and $\omega_2$, and with OAM modes $l = \pm l_1$ and $l = \pm l_2$, respectively.

The vortex beam's electric field is

$$E(r,\theta,t) = B_1(r)\exp(-\mathrm{i}\omega_1 t)\exp(il_1\theta) + B_1(r)\exp(-\mathrm{i}\omega_1 t)\exp(-il_1\theta)$$
$$+ B_2(r)\exp(-\mathrm{i}\omega_2 t)\exp(il_2\theta) + B_2(r)\exp(-\mathrm{i}\omega_2 t)\exp(-il_2\theta)\,, \quad (S1)$$

where $r$ is the radial coordinate of the beam, $\theta$ is the azimuthal angle, and $t$ is time. When this field is incident on a spinning surface whose reflectivity is spatially inhomogeneous, the reflected field is composed of OAM modes characterized by topological charge $m$:

$$\sum_m \left[ B_1(r)\,A_{m-l_1}(r)\exp(-\mathrm{i}\omega_1 t)\exp(-\mathrm{i}[m-l_1]\Omega t)\exp(\mathrm{i}m\theta) \right.$$
$$+ B_1(r)\,A_{m+l_1}(r)\exp(-\mathrm{i}\omega_1 t)\exp(-\mathrm{i}[m+l_1]\Omega t)\exp(\mathrm{i}m\theta)$$
$$+ B_2(r)\,A_{m-l_2}(r)\exp(-\mathrm{i}\omega_2 t)\exp(-\mathrm{i}[m-l_2]\Omega t)\exp(\mathrm{i}m\theta)$$
$$\left. + B_2(r)\,A_{m+l_2}(r)\exp(-\mathrm{i}\omega_2 t)\exp(-\mathrm{i}[m+l_2]\Omega t)\exp(\mathrm{i}m\theta) \right]\,. \quad (S2)$$

Here, $B(r)$ and $A(r)$ are the complex amplitudes of the incident and reflected fields, respectively, and $\Omega$ is the angular frequency of the rotating target. The square of this electric field is proportional to its intensity. Among various terms, there will be cross terms of the following type:

$$\sum_{m \neq n} |B_1(r)|^2 \left| A_{m-l_1}(r)A_{n-l_1}(r)\right| \exp(-\mathrm{i}[m-n]\Omega t)\exp(\mathrm{i}[m-n]\theta) + c.c$$
$$= \sum_{m<n} 2|B_1(r)|^2 \left| A_{m-l_1}(r)A_{n-l_1}(r)\right| \cos([m-n]\Omega t - [m-n]\theta)\,,$$

or, *e.g.*, of the type:

$$\sum_{m \neq n} \left| B_1(r)B_2(r)A_{m-l_1}(r)A_{n-l_2}(r)\right| \exp(-\mathrm{i}\Delta\omega t)\exp(-\mathrm{i}[(m-n)-(l_1-l_2)]\Omega t)\exp(\mathrm{i}[m-n]\theta) + c.c$$
$$= \sum_{m \neq n} 2|B_1(r)B_2(r)A_{m-l_1}(r)A_{n-l_2}(r)| \cos(\Delta\omega t + [(m-n)+(l_1-l_2)]\Omega t - [m-n]\theta)\,,$$

where $\Delta\omega = \omega_2 - \omega_1$, and where for simplicity we have ignored any additional phase terms of the type, *e.g.*, $\phi(r) = \mathrm{angle}\big[B_1(r)A_{m-l_1}(r)B_2^*(r)A_{n-l_2}^*(r)\big]$ [1]. As a photodetector (with radial extent $R$) would see the integral $\int_0^R \int_0^{2\pi} I(r,\theta,t)\,r\mathrm{d}\theta\mathrm{d}r$, all above-like cross terms will disappear as $\int_0^{2\pi}\cos([m-n]\theta)\,\mathrm{d}\theta = 0$ for all $m \neq n$. This justifies the main text's claim that there are common detection modes $m = n$. We can therefore omit the cross terms and write the intensity as

$$I(r,t) = \sum_m \left|B_1(r)A_{m-l_1}(r)\right|^2 + \sum_m \left|B_1(r)A_{m+l_1}(r)\right|^2 + \sum_m \left|B_2(r)A_{m-l_2}(r)\right|^2$$

$$+ \sum_m \left|B_2(r)A_{m+l_2}(r)\right|^2 + \sum_m |B_1(r)|^2 \left|A_{m-l_1}(r)A_{m+l_1}(r)\right| \exp(\mathrm{i}2l_1\Omega t)$$

$$+ \sum_m |B_2(r)|^2 \left|A_{m-l_2}(r)A_{m+l_2}(r)\right| \exp(\mathrm{i}2l_2\Omega t)$$

$$+ \sum_m \left|B_1(r)B_2(r)A_{m-l_1}(r)A_{m-l_2}\right| \exp(-\mathrm{i}\Delta\omega t)\exp(\mathrm{i}[l_1 - l_2]\Omega t)$$

$$+ \sum_m \left|B_1(r)B_2(r)A_{m-l_1}(r)A_{m+l_2}\right| \exp(-\mathrm{i}\Delta\omega t)\exp(\mathrm{i}[l_1 + l_2]\Omega t)$$

$$+ \sum_m \left|B_1(r)B_2(r)A_{m+l_1}(r)A_{m-l_2}\right| \exp(-\mathrm{i}\Delta\omega t)\exp(-\mathrm{i}[l_1 + l_2]\Omega t)$$

$$+ \sum_m \left|B_1(r)B_2(r)A_{m+l_1}(r)A_{m+l_2}\right| \exp(-\mathrm{i}\Delta\omega t)\exp(-i[l_1 - l_2]\Omega t) + c.c.$$

$$= \sum_m \left|B_1(r)A_{m-l_1}(r)\right|^2 + \sum_m \left|B_1(r)A_{m+l_1}(r)\right|^2 + \sum_m \left|B_2(r)A_{m-l_2}(r)\right|^2$$

$$+ \sum_m \left|B_2(r)A_{m+l_2}(r)\right|^2 + 2\sum_m |B_1(r)|^2 \left|A_{m-l_1}(r)A_{m+l_1}(r)\right| \cos(2l_1\Omega t)$$

$$+ 2\sum_m |B_2(r)|^2 \left|A_{m-l_2}(r)A_{m+l_2}(r)\right| \cos(2l_2\Omega t)$$

$$+ 2\sum_m \left|B_1(r)B_2(r)A_{m\mp l_1}(r)A_{m\mp l_2}\right| \cos(\Delta\omega t \pm [l_1 - l_2]\Omega t)$$

$$+ 2\sum_m \left|B_1(r)B_2(r)A_{m\mp l_1}(r)A_{m\pm l_2}\right| \cos(\Delta\omega t \pm [l_1 + l_2]\Omega t) \,. \tag{S3}$$

All in all, the signal consists of a DC component and intensity modulation at angular frequencies $2l_1\Omega$ and $2l_2\Omega$, as expected. In addition, there will be cross-coupling between the different OAM modes leading to additional modulation at $[l_1 \pm l_2]\Omega$. However, these will appear at high frequencies as sidebands around the beat notes $\Delta\omega = \omega_1 - \omega_2$ and can be easily filtered out.

## 2. Signal construction in the emulated vortex-comb spectroscopy measurement

To arrive at the spectrum shown in Fig. 4 (c) of the main text, a Voigt line profile with a constant offset and a sine function (to consider a possible etalon) was fitted to each of the spectra measured with the separate OAM modes (*i.e.*, to spectra such as the ones shown in panels a and b of Fig. 4 of the main text) with the Doppler half-width-at-half-maximum (HWHM) fixed to its theoretical value. The fitted background was subtracted, and the spectrum was normalized with the height of the fitted Voigt profile. The normalization step corresponds to a calibration measurement where the laser is parked on top of the absorption feature to consider both the OAM-dependent intensity modulation amplitude and the response of the CEPAS detector at each modulation frequency (see Supplement 1 of Ref. [2]). We note that also in a more conventional transmission spectrum measurement (*i.e.*, in a measurement without CEPAS detection) one would typically need to measure a separate background spectrum for normalization. The background subtraction is needed because in CEPAS measurements the modulation peaks typically lie on top of a noise floor (see Supplementary information of Ref. [3]), and because a constant absorption offset can emerge if any of the incident optical power is absorbed at the cell windows or walls. The constant absorption offset can also have OAM

dependence as the diameter of the beam increases as a function of the OAM mode [4]; in the measurements reported here, the petal pattern diameter on the rotating mirror was 1.7 mm for the OAM 1 and 4.1 mm for the OAM 11 mode.

}
\end{document}